\documentstyle[aps,prl,twocolumn,epsf]{revtex}
\draft
\begin{document}
\title{
How long does it take for the Kondo effect to develop?}
\author{Peter Nordlander}
\address{
Department of Physics and Rice Quantum Institute,
Rice University, Houston, Texas 77251-1892 }
\author{Michael Pustilnik and Yigal Meir}
\address{Physics Department, Ben Gurion University, Beer
Sheva, 84105, Israel}
\author{Ned S. Wingreen}
\address{NEC Research Institute, 4 Independence Way,
Princeton, NJ 08540}
\author{David C. Langreth}
\address{Department of Physics and Astronomy,
Rutgers University, Piscataway, NJ 08854-8019
}

\maketitle
\thispagestyle{empty}
\begin{abstract}
\widetext
The time-development of the Kondo effect is theoretically
investigated by studying
a quantum dot suddenly shifted into the Kondo regime by
a change of voltage on a nearby gate. Using time-dependent versions
of both the
Anderson and Kondo Hamiltonians, it is shown that
after a time $t$ following the voltage shift, the form of the Kondo
resonance matches the {\it time-independent} resonance
at an effective temperature $T_{\rm  eff} = T/\tanh(\pi T t/2)$. Relevance of
the buildup of the Kondo resonance to the
transport current through a quantum dot is demonstrated.
\end{abstract}

\narrowtext

The Kondo effect in quantum dots has been observed
in several recent experiments\cite{kastner97}.
Beyond verifying theoretical predictions \cite{glazman89,goodguys}, these
experiments demonstrate that quantum dots can serve as an important new tool
to study strongly correlated electron systems. Unlike magnetic impurities
in metals,  the physical parameters of the quantum dot can be varied
continuously,  which allows,  for example,  systematic experimental study
of the crossover between the Kondo,  the mixed-valence, and the non-Kondo
regimes. Moreover, the quantum dot
system opens the possibility of directly observing the
time-dependent response of a Kondo system, as
there is a well developed technology for
applying time-dependent perturbations to
dots\cite{acreview}. Along these lines,
several theoretical works have addressed the
behavior of a Kondo impurity subject to ac driving
\cite{drivenkondo}.
However, a clearer picture of the temporal development
of many-body correlations is obtained if
the impurity is subject to a sudden shift in energy.
Specifically, by applying a step-like impulse to a
nearby gate, the dot can be suddenly shifted
into the Kondo regime, and the buildup of the
correlated state observed in the transport current.

In this Letter, we analyze the behavior of
a quantum dot following a sudden shift into
the Kondo regime. The time-dependent
spectral function is evaluated within the non-crossing approximation
(NCA)\cite{goodguys,bickers,LangrethNordlander91PRB,ShaoetAl194PRB},
as is the transport current in response to a pulse train.
The latter provides an experimental window on the
development of the Kondo resonance. Employing the Kondo Hamiltonian,
we show that a finite development time $t$ is perturbatively
equivalent to
an increase in the effective
\pagebreak\vspace*{1.4in}
temperature.

\begin{figure}[h]
\centerline{\epsfxsize=0.45\textwidth
\epsfbox{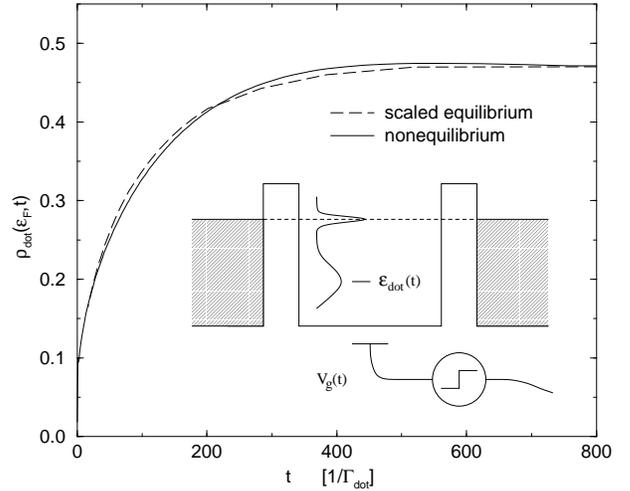}
}
\caption{Solid curve: time-dependent spectral density
$\rho_{\rm dot}(\epsilon_F,t)$
at the Fermi energy vs. time $t$ following the same step-function
change in the level energy used in Fig.~1. The temperature is
$T=0.0025$.
Dashed curve: equilibrium spectral density $\rho_{\rm dot}(\epsilon_F)$
at the Fermi energy with temperature set according to
Eq.~(7):
$T_{\rm eff}(t) = T/\tanh(\pi T t/2)$, with $T = 0.0025$.
Inset: schematic of the quantum-dot single electron transistor.}
\label{fig:rhovst}
\end{figure}

We treat a quantum dot coupled by tunnel barriers to two leads (inset to
Fig.~\ref{fig:rhovst}). Only one
spin-degenerate level on the dot is considered, which is a good approximation
at low temperatures. A time-dependent voltage $V_g(t)$ is
applied to a nearby gate, causing a proportionate shift in the energy of the
level
$\epsilon_{\rm dot}(t)$.
If the Coulomb interaction between electrons prevents double occupancy
of the dot, the system  is described by the $U = \infty$ Anderson
Hamiltonian for a magnetic impurity,
\begin{equation}
\sum_\sigma \!
\epsilon_{\rm dot}(t)\,
n_\sigma +\sum_{k\sigma}\!\left[\epsilon_{k\sigma}n_{k\sigma}
+(V_k c^\dagger_{k\sigma}c_\sigma + {\rm H.c.})\right],
\label{hamiltonian}
\end{equation}
with the constraint that the occupation of the dot cannot
exceed one electron.
Here $c^\dagger_\sigma$ creates an electron of spin $\sigma$
in the quantum dot, with $n_\sigma$ the corresponding
number operator; $c^\dagger_{k\sigma}$ creates an electron in the leads,
with $k$ representing all quantum numbers other than spin,
including the labels, left and right, for the leads.
$V_k$ is the tunneling matrix element
through the appropriate barrier.
The quantum dot is occupied by a single electron provided the level
energy $\epsilon_{\rm dot}$ lies at least a resonance width
$\Gamma_{\rm dot}$ \cite{Gnote}
below the chemical potential of the leads. At low temperatures,
the resulting free spin on the dot forms a singlet with a
spin drawn from the electrons in the leads -- this is the Kondo
effect. The Kondo temperature, beneath which the strongly correlated
state is established, is given by
$T_K \simeq D' \exp(-\pi|\epsilon_{\rm dot}|/\Gamma_{\rm dot})$,
where $D'$ is a high energy cutoff \cite{dprime}.
The signature of this correlated state is
a peak at the Fermi energy in the spectral density of the dot
electrons. This peak, in turn, dramatically enhances transport through
the dot, allowing perfect transmission \newpage
\begin{figure}[h]
\centerline{\epsfxsize=0.45\textwidth
\epsfbox{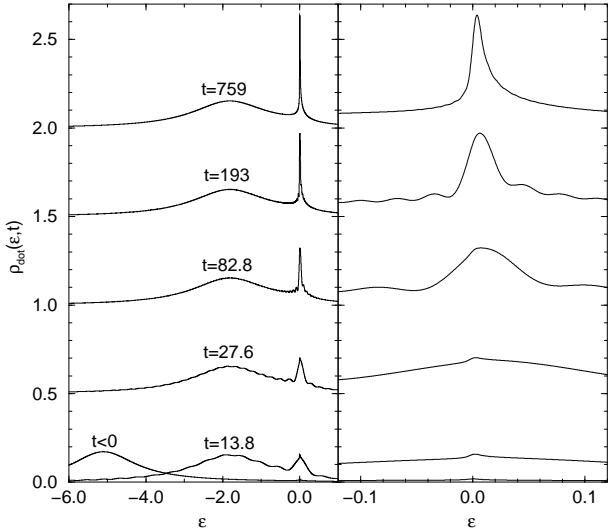}
}
\caption{Spectral density $\rho_{\rm dot}(\epsilon,t)$ vs.\ energy $\epsilon$
at various times following a step-function change in the level energy
$\epsilon_{\rm dot}(t) = -5 + 3 \theta(t)$. The ordinates for positive
times are successively offset by 0.5 units.
For $t<0$, $\rho_{\rm dot}
(\epsilon,t)$ is identical
to the equilibrium spectral density at $\epsilon_{\rm dot} = -5$
while for the largest time shown it is indistinguishable on this scale from
the equilibrium spectral density at $\epsilon_{\rm dot} = -2$.
Throughout this work energies are given
in units of $\Gamma_{\rm dot}$, and times in units
of $1/\Gamma_{\rm dot}$, with $\hbar=1$. Here $T=0.0025$.}
\label{fig:dos}
\end{figure}
\noindent at zero temperature\cite{glazman89}.

We employ the non-crossing approximation (NCA) to analyze the
spectral density and transport through the dot in the presence
of a {\it time-dependent} level energy $\epsilon_{\rm dot}(t)$.
The NCA is based on an exact transformation of the $U = \infty$ Anderson
model in Eq.~(\ref{hamiltonian}) into a slave-boson Hamiltonian
\cite{bickers}. The latter is then solved self-consistently
to second order in the tunneling matrix elements $V_k$.
The NCA approximation gives reliable results for temperatures
down to $T < T_K$, and its time-dependent formulation has
been discussed at length in previous
works\cite{LangrethNordlander91PRB,ShaoetAl194PRB}.
We define a time-dependent spectral density
for the dot electrons as\cite{jauho94}
\begin{equation}
\rho_{\rm dot}(\epsilon,t)\equiv {\rm Re}\int_{0}^{\infty}\frac{d\tau}{\pi}
e^{i\epsilon\tau} \langle
\{c_\sigma(t),c_\sigma^\dagger(t-\tau)\}\rangle.
\end{equation}
The spectral density is the same for both spins in the absence
of a magnetic field.
In equilibrium, $\rho_{\rm dot}(\epsilon,t)$ is independent of
$t$ and reduces to the usual
density of states. Furthermore, $\rho_{\rm dot}(\epsilon,t)$ is
causal, {\it i.e.} it depends only on the behavior of the level
energy for times earlier than $t$. In Fig.~(\ref{fig:dos}), we have
plotted the time-dependent spectral density for several times
following an abrupt shift of the level energy \cite{abrupt}. Before the shift,
the level energy is so low, $\epsilon_{\rm dot} = -5$,
that the Kondo temperature is much
smaller than the physical temperature, and so there is no noticeable
Kondo peak in the spectral density. At $t=0$, the level energy is
abruptly shifted to $\epsilon_{\rm dot} = -2$, giving $T_K \sim 10^{-3}$,
comparable to T. The Kondo peak  \pagebreak
\noindent thereafter grows with a characteristic
time dependence shown in Fig.~(\ref{fig:rhovst}), approaching a new
equilibrium value at long times $t \sim 1/T_K$.

To develop an analytical theory for the time development
of the Kondo resonance, we
consider the time-dependent Kondo Hamiltonian,
\begin{equation}
H_{K}=\sum_{k\sigma }\!\epsilon _{k\sigma }n_{k\sigma }
+J(t)\,{\bf S}\cdot {\bf %
s}  \label{HK}
\end{equation}
\[
{\bf s}=\sum_{\alpha \alpha ^{\prime }}\psi _{\alpha }^{\dagger }\frac{{\bf %
\sigma }_{\alpha \alpha ^{\prime }}}{2}\psi _{\alpha ^{\prime }},\;{\bf S}%
=\sum_{\beta \beta ^{\prime }}c_{\beta }^{\dagger }\frac{{\bf \sigma }%
_{\beta \beta ^{\prime }}}{2}c_{\beta ^{\prime }},\;n_{c}=\sum_{\beta
}c_{\beta }^{\dagger }c_{\beta }.
\]
Here $\psi _{\alpha }^{\dagger }$
creates a conduction-band electron at the site of the Kondo impurity, and
the ${\bf \sigma }$ are the Pauli spin matrices. For near Fermi surface
properties, the Anderson Hamiltonian reduces to the simpler
Kondo Hamiltonian with $%
J(t)=2|V_{k_{F}}^{2}/\epsilon _{{\rm dot}}(t)|$ when the site is occupied by a
single electron \cite{SchriefferWolff}. For the case of interest, in
which the level energy $\epsilon_{\rm dot}$ is suddenly shifted into
the Kondo regime at $t = 0$, it is adequate to consider a sudden switching on
of the Kondo coupling, $J(t) = J \theta(t)$. We have written the Kondo
Hamiltonian in terms of Abrikosov's
pseudofermion representation \cite{abrikosov}, where $c_{\beta }^{\dagger }$
creates a pseudofermion of spin $\beta $, which represents the magnetic
impurity. In order to restrict the states to the physical subspace, we have
to impose a constraint $n_{c}=1$. This can be done by adding a term $\lambda
n_{c}$ to the Hamiltonian \cite{abrikosov} and taking the limit $\lambda
\rightarrow \infty $. Since we are interested in a sudden switching on of
the Kondo coupling $J\,{\bf S}\cdot {\bf s}$, we are able
to implement the constraint $n_c =1$
by explicitly creating a
pseudofermion at $t=0$. Then, as the \pagebreak
\begin{figure}[h]
\centerline{\epsfxsize=0.30\textwidth
\epsfbox{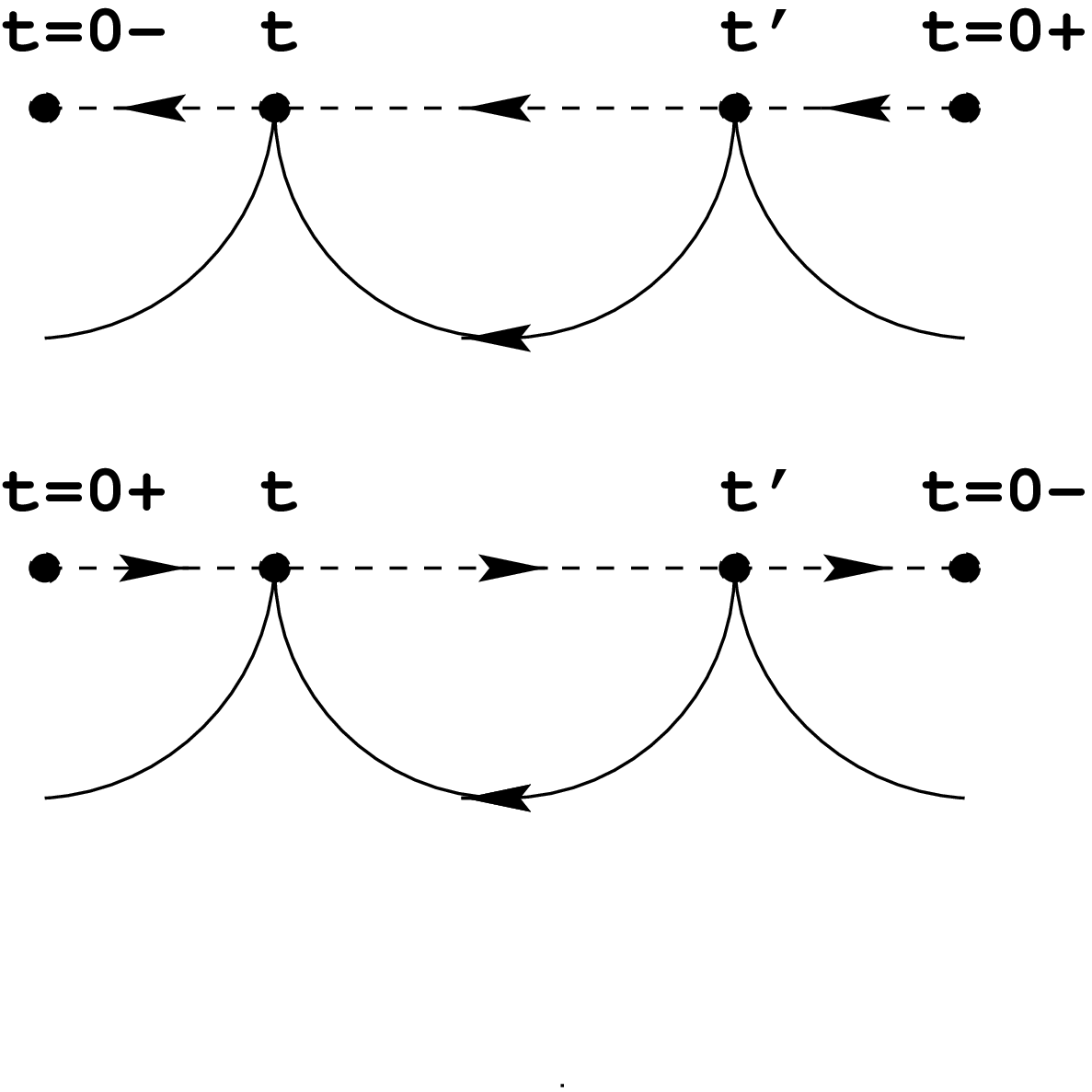}
}
\caption{Contributions of order $J^2$ to the renormalized conduction
electron scattering vertex, from the Kondo Hamiltonian in Eq.~(3).
Solid lines are conduction electron propagators and dashed lines are
pseudofermion propagators. Summation over internal spins is implied.}
\label{fig:vertices}
\end{figure}
\noindent pseudofermion number
is conserved by $H_{K}+\lambda n_{c}$, and we have
$n_c=0$ for $t<0$ because of the large
pseudofermion
 energy $\lambda \rightarrow \infty $, we obtain an
abrupt turn on of the Kondo coupling at $t=0$ and all later
expectations are taken in the physical subspace $n_{c}=1$.

The analytical signature of the Kondo effect is the logarithmic
divergence of perturbation theory in the dimensionless coupling $J\rho$,
where $\rho$ is the density of conduction electron states per spin
direction at the Fermi level.
Indeed, for $T < T_K$ perturbation theory in $J\rho$ fails, even for
small $J\rho$. For $T > T_K$,
temperature cuts off the logarithmic divergences and perturbation
theory is reliable \cite{abrikosov}. We find that a finite time $t$ following a
sudden
switching on of the Kondo coupling also results in a convergent
perturbation theory. To demonstrate this,
we focus on the simplest quantity that diverges in
perturbation theory. Specifically, we calculate the scattering
vertex $\gamma ^{pp}(t,t')$ to order $J^2$.
Physically, this quantity represents the lowest order change
in $J$ due to multiple scattering from the Kondo impurity.
{Since abruptly turning on the Kondo
coupling creates a nonequilibrium state of the system, we use Keldysh Green
functions with }$p=\pm 1$ for the outward/backward branches{. In time,
the Keldysh
contour runs from }$-\infty $ to $\infty$ ($p=+1$)
and then from $\infty$ to  $-\infty
$ ($p=-1$). As shown in Fig.~(\ref{fig:vertices}), there are two
contributions at order $J^{2}$, one with the conduction electron line and
the pseudofermion line parallel and one with the lines antiparallel.
Evaluating the diagrams in Fig.~({\ref{fig:vertices}), and keeping only
logarithmically divergent contributions in addition to the bare vertex, we find
\begin{eqnarray}
\gamma ^{pp'}(t,t')&=&p\delta _{pp^{\prime }}\frac{J}{4}\left( {\bf %
\sigma }_{\alpha \alpha ^{\prime }}{\bf \sigma }_{\beta \beta ^{\prime
}}\right) \theta \left( t\right) \theta \left( t'\right) \nonumber \\
&&\times\left[ 1-%
\frac{J}{2}G_{0}^{pp}\left( t-t'\right)
\mathop{\rm sgn}%
\left( t-t'\right) \right].   \label{gamma}
\end{eqnarray}
(Note that\ in this order there is no logarithmic contribution
that is off-diagonal in the Keldysh indices.)
Here $G_{0}^{pp}\left( t-t'\right) $ is the {bare time-ordered (for }$%
p=+1${) or anti-time-ordered (for }$p=-1$){\ Green function for conduction
electrons at the site of the Kondo impurity. For  }$|t-t'| \gg 1/D$ ($D$
is a high-energy cutoff) it takes the form\cite{Yuval}
\begin{equation}
{G_{0}^{pp}(t-t')\rightarrow {\frac{{-\pi \rho T}}{\sinh [\pi
T(t-t')]}}.}
\end{equation}
{Fourier transforming (\ref{gamma}) with respect to the time
difference $t-t'$, and taking the limit of zero frequency
to obtain the effective scattering vertex at time $t$ for electrons
near the Fermi energy, we find
\begin{equation}
\gamma \left( t,\omega \rightarrow 0\right) \propto J\left\{ 1+
\frac{1}{2}\rho J\ln
\left[ \frac{D}{T}\tanh
\left( \frac{\pi Tt}{2}\right) \right] \right\}.
\end{equation}
For }$Tt\gg 1$ this reduces to the usual equilibrium form, $\gamma \propto J%
\left[ 1+\case{1}{2}\rho J\ln \frac{D}{T}\right] $,
with the logarithmic divergence cut off
only by temperature. However, since in our case the Kondo coupling exists
only for times $t>0$, the result contains an additional cutoff due to the
finite time allowed for spin-flip scattering.
Formally, the finite time $t$ since the onset of the Kondo coupling
can be absorbed into an increase in the effective temperature,
\begin{equation}
T_{\rm  eff} = {{T} \over {\tanh(\pi T t/2)}  }.
\label{teff}
\end{equation}

How accurately does this effective temperature represent the time
development of the Kondo resonance at the Fermi surface? To test
the applicability of Eq.~(\ref{teff}) beyond perturbation theory,
we have compared the time-dependent NCA results to
{\it time-independent equilibrium}
NCA results at the corresponding effective temperature.
The agreement, with no free parameters, is quite good as is seen in
Fig.~({\ref{fig:rhovst}).
Note that at short times
$T_{\rm  eff} \simeq 2/\pi t$. Hence, the buildup
of the Kondo resonance is governed by a type of energy-time uncertainty
relation: after a time $t$ the Kondo resonance is cut off by an
energy $\sim 1/t$ \cite{Jcubed}.
Thus we expect saturation of the Kondo peak at a time
$t\sim 1/T_K$,  as indeed is observed numerically.

Finally, we consider how the buildup of the Kondo resonance can
be observed experimentally. In steady state, the linear-response
conductance $G$ through a dot symmetrically coupled to its leads is given
by\cite{meir92}
\begin{equation}
G = {e^2 \over \hbar} {{\Gamma_{\rm dot} } \over {2} }
\int \!d\epsilon\,\rho_{\rm dot} (\epsilon)\left(-\frac{\partial
f(\epsilon)}{\partial\epsilon}\right),
\label{conductance}
\end{equation}
where $f(\epsilon)$ is the Fermi function, and $\hbar$ is explicitly
included for clarity. If a periodic gate
voltage is applied to the dot, formula~(\ref{conductance}) is
still valid if $G$ is replaced by the {\it time-averaged}
conductance $\langle G \rangle$, and
$\rho_{\rm dot}(\epsilon)$ is replaced by the average
of the time-dependent spectral density
$\langle\rho_{\rm dot}(\epsilon,t)\rangle$.
Consider a periodic signal consisting of an ``on''
pulse of duration $\tau_{\rm on}$
which brings the dot into the
Kondo regime followed by an ``off'' pulse
which moves it back out of the Kondo regime.
During each on pulse, $\rho_{\rm dot}(\epsilon_F,t)$ will build up to a maximum
at time $\tau_{\rm on}$ and then rapidly decrease back to a low value
during the off pulse.
The differential increase of conductance as the
duration of the on \pagebreak
\begin{figure}[h]
\centerline{\epsfxsize=0.45\textwidth
\epsfbox{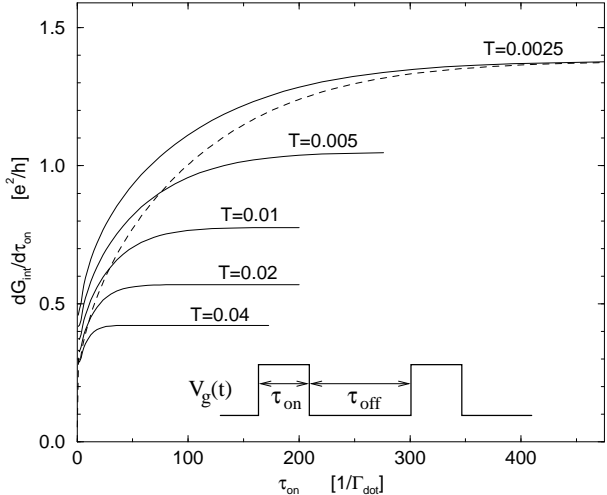}
}
\caption{Solid curves: derivative of $G_{\rm int}$ (in units of
$e^2/h$) with respect to duration
$\tau_{\rm on}$ of ``on" gate-voltage pulses, at various temperatures.
$G_{\rm int}$ is the conductance integrated over a full cycle
of gate voltage. Dashed curve:
$-\pi \int \! d\epsilon\, \Gamma_{\rm dot}\, f'(\epsilon)\rho_{\rm dot}
(\epsilon,t=\tau_{\rm on})$ for $T=0.0025$.
Inset: schematic periodic
gate-voltage pulse train. The level energy is
$\epsilon_{\rm dot} = -2$ in the on state and $\epsilon_{\rm dot} = -5$
in the off state.
The duration of the off period, $\tau_{\rm off}$ is long enough to allow
transients from each on pulse to die out.}
\label{fig:dgdton}
\end{figure}
\noindent pulse is increased
will therefore reflect
 the magnitude of the
spectral density near or at the Fermi energy
at a time $\tau_{\rm on}$ following the shift into the
Kondo regime. In Fig.~(\ref{fig:dgdton}), we have plotted the differential
with respect to $\tau_{\rm on}$ of the conductance, with a fixed off-pulse
duration $\tau_{\rm off}$. The conductance is integrated over the period,
rather than time-averaged, to remove effects due to the changing duration
of the period, {\it i.e.}
$G_{\rm int} = (\tau_{\rm on} + \tau_{\rm off})\, \langle G \rangle$.
This measurable transport quantity provides a  probe of the
time-development of the Kondo resonance \cite{decay}.

In conclusion, we have analyzed the response of a quantum
dot to a sudden shift of gate voltage which takes the dot into the regime
of the Kondo effect.  The buildup of many-body correlations
between the dot and the leads  follows an uncertainty principle:
at time $t$
the Kondo resonance is cut off by an energy $\sim 1/t$.
Within perturbation theory in the Kondo coupling, we find that the
finite time $t$ plays the role of an increased effective temperature
$T_{\rm  eff} = T/\tanh(\pi T t/2)$. To experimentally probe the buildup
of the Kondo resonance, we propose applying a train of square gate-voltage
pulses to the dot. The derivative of current with respect
to duration of the ``on" pulse accurately reproduces the time-dependent
amplitude of the Kondo resonance.

The work was supported in part by NSF grants DMR 95-21444 (Rice) and
 DMR 97-08499 (Rutgers). Work
at BGU was supported by the The Israel Science Foundation -
Centers of Excellence Program. One of us (MP) acknowledges the
support of a Kreitman Fellowship.


\begin{thebibliography}{10}
\bibitem{Gnote}
We define $\Gamma_{\rm
  dot}(\epsilon) = 2
  \pi \sum_k |V_k|^2 \delta(\epsilon -\epsilon_k)$, a slowly varying quantity.
 The
  notation $\Gamma_{\rm dot}$ with no energy specified will always refer the
  value at the Fermi level.
\bibitem{dprime} For $U=\infty$, $D' \simeq \sqrt{D\Gamma_{\rm dot}/4}$,
where $2D$ is the effective bandwidth.  The calculations here used
a parabolic band of total width $40\Gamma_{\rm dot}$.
\bibitem{abrupt}
Our calculations are based on the approximation that the switching time,
$\tau_s$, is exactly zero. In reality, $\tau_s$ is always a finite time. Our
results
are valid for finite $\tau_s$ as well, provided that $t\gg\tau_s$.
\bibitem{Jcubed}
By evaluating the conduction electron self-energy to order
$J^3$, we have directly confirmed the $\sim 1/t$ cutoff for
the Kondo peak in the spectral density.
\end{thebibliography}
\end{document}